\documentclass[12pt]{article}
\usepackage{amsmath,amssymb}
\newcommand{\be}{\begin{equation}}
\newcommand{\ee}{\end{equation}}
\newcommand{\bea}{\begin{eqnarray}}
\newcommand{\eea}{\end{eqnarray}}

\def\P{Poincar\'e }

\begin{document}
\renewcommand {\theequation}{\thesection.\arabic{equation}}
\renewcommand {\thefootnote}{\fnsymbol{footnote}}
\vskip1cm
\begin{flushright}
\end{flushright}
\vskip1cm
\begin{center}
{\large\bf Twist as a Symmetry Principle\\ and the Noncommutative
Gauge Theory Formulation}

\vskip .7cm {\bf{{M. Chaichian$^{a}$, A. Tureanu$^{a}$ and G.
Zet$^{a,b}$}}

{\it $^{a}$High Energy Physics Division, Department of Physical
Sciences,
University of Helsinki\\
\ \ {and}\\
\ \ Helsinki Institute of Physics,\\ P.O. Box 64, FIN-00014
Helsinki, Finland\\
$^{b}$Department of Physics, "Gh. Asachi" Technical University,\\Bd.
D. Mangeron 67, 700050 Iasi, Romania }}
\end{center}
\vskip1cm
\begin{abstract}

Based on the analysis of the most natural and general ansatz, we
conclude that the concept of twist symmetry, originally obtained for
the noncommutative space-time, cannot be extended to include
internal gauge symmetry. The case is reminiscent of the
Coleman-Mandula theorem. Invoking the supersymmetry may reverse the
situation.

\end{abstract}

\vskip1cm

\newpage

\section{Introduction}

For field theories on the noncommutative space-time with
Heisenberg-like commutation relation
\be\label{cr}[\hat x_\mu,\hat x_\nu]=i\theta_{\mu\nu}\,,\ee
where $\theta_{\mu\nu}$ is an antisymmetric matrix, the traditional
framework has been the Weyl-Moyal correspondence, by which to each
field operator $\Phi(\hat x)$ corresponds a Weyl symbol $\Phi(x)$,
defined on the commutative counterpart of the space-time. An
essential aspect of this correspondence is that, in the action
functional, the products of field operators, e.g. $\Phi(\hat
x)\Psi(\hat x)$ is replaced by the Moyal $\star$-product of Weyl
symbols, $\Phi(x)\star\Psi(x)$, where
\be\label{star}\star=\exp{\left(\frac{i}{2}\theta_{\mu\nu}\overleftarrow\partial^\mu\overrightarrow\partial^\nu\right)}\
.\ee
In this correspondence, the operator commutation relation (\ref{cr})
becomes
\be\label{star cr}[ x_\mu, x_\nu]_\star=x_\mu\star x_\nu-x_\nu\star
x_\mu=i\theta_{\mu\nu}\ee
and noncommutative models have been built by simply taking their
commutative counterparts and replacing the usual multiplication by
$\star$-product (see \cite{SW} and references therein).

It turns out that such noncommutative models, although they lack
Lorentz symmetry, are invariant under the twisted \P algebra
\cite{CKT}, deformed with the Abelian twist element
\be\label{abelian twist}{\cal
F}=\exp\left({\frac{i}{2}\theta^{\mu\nu}P_\mu\otimes
P_\nu}\right),\ee
where $P_\mu$ are the generators of translations. The twist induces
on the algebra of representation of the \P algebra the deformed
multiplication
\be\label{twist prod} m\circ(\phi\otimes\psi)=\phi\psi\rightarrow
m_\star\circ(\phi\otimes\psi)=m\circ{\cal
F}^{-1}(\phi\otimes\psi)\equiv \phi\star\psi\,,\ee
which is nothing else but the $\star$-product (\ref{star}).
Important consequences for the representation theory of the
noncommutative fields arise from here.

In parallel with the NC QFT models, NC gauge theories have been
constructed using the same prescription, of taking the Lagrangian of
the commutative theory and replacing the usual multiplication by the
$\star$-product (\ref{star}) \cite{Hayakawa}. By construction, such
theories are twisted-\P invariant, if we use the twist element
(\ref{abelian twist}). However, as far as the gauge invariance is
concerned, the models are invariant under $\star$-gauge
transformations. For example, in the case of the gauge $U_\star(n)$
group, an arbitrary element of the group will be
\be U(x)=\exp_\star({i\alpha^a(x)T_a}) \ee
where $T_a$, $a=1,...,n^2$ are the generators of the $U(n)$ group,
with the algebra $[T_a,T_b]=if_{abc}T_c$, $\alpha^a(x)$,
$a=1,...,n^2$ are the gauge parameters and the $\star$-exponential
means
\be
\exp_\star({i\alpha^a(x)T_a})=1+i\alpha^a(x)T_a+\frac{1}{2!}(i)^2\alpha^a(x)\star\alpha^b(x)T_aT_b+...\ee
The use of the $\star$-product in the formulation of gauge theories
imposes strict constraints on the noncommutative gauge symmetry,
among which is the fact that only NC gauge $U(n)$ groups close (and
not $SU(n)$). Moreover, there is a no-go theorem \cite{nogo} stating
that only certain representations of the gauge group are allowed
(fundamental, antifundamental and adjoint) (see also
\cite{terashima}) and the matter fields can be charged under at most
two gauge groups. We have to emphasize that although these gauge
theories are twisted-\P invariant, the $\star$-gauge transformations
are implemented separately, in the sense that the coproduct of the
gauge generators is not twisted with the Abelian twist (\ref{abelian
twist}).

Recently, an attempt was made to twist also the gauge algebra, i.e.
to extend the \P algebra by the gauge algebra, as semidirect
product, and to twist the coproduct of the gauge generators with the
same Abelian twist (\ref{abelian twist}) \cite{Vassilevich,Wess}.
The result seemed to be spectacular: the same theories, which
previously were shown to be subject to the no-go theorem \cite{nogo,
terashima}, were now claimed to be invariant under any usual (not
noncommutative) gauge group and to admit any representations, just
as in the commutative case. The latter approach was shown
\cite{twist gauge} however to be in conflict with the very idea of
gauge transformations, since it assumed implicitly that if a field
is transformed according to a given representation of the gauge
algebra, then its derivatives of any order also transform according
to the representations of the gauge algebra, which is obviously not
the case.

The question arises whether the concept of {\it twist} appears as a
{\it symmetry principle} in constructing NC field theories: any
symmetry that such theories may enjoy, be it space-time or internal
symmetry, global or local, should be formulated as a twisted
symmetry. In pursuit of this idea, in this letter we take the most
general ansatz for a non-Abelian twist, which, in the absence of the
gauge interaction, reduces to the Abelian twist (\ref{abelian
twist}). We shall show that the twisting of the gauge
transformations is not possible, in a manner compatible with the
representations of the gauge algebra and keeping at the same time
the Moyal space defined by (\ref{star cr}) as underlying space of
the theory.

\section{Necessity of a symmetry principle for noncommutative field
theories}

The necessity of a new approach to noncommutative gauge theories
arises both from internal gauge symmetries and the gravitational
theory.

{\it Noncommutative internal gauge symmetry}

The essential physical implication of the twisted \P symmetry is
that the representation content of this quantum symmetry and of
usual \P symmetry are the same. As a consequence, the asymptotic
fields are the same in commutative and noncommutative field
theories. This legitimates the perturbative approach to NC QFT,
starting from the representation content of \P algebra (for details,
see \cite{CKT}).

On the other hand, any application such as model building has to
circumvent one way or another the no-go theorem
\cite{nogo,terashima}. The ways for by-passing the restrictions
imposed by the no-go theorem (e.g. by dressing the fields with
Wilson lines or by invoking enveloping algebra-valued fields) are
not unique and lack justification. A twisted symmetry principle
would provide a truly solid base for the formulation of
noncommutative gauge theories.

{\it Noncommutative gravitational theory}

 NC gravitational effects
have been recently calculated \cite{LAG-grav} from string theory
with antisymmetric background field, i.e. in the same theory as the
one which gave rise in the low-energy limit to the usual
noncommutative field theories \cite{SW}. It turns out that, in the
case of NC gravitational interactions, string theory contains a much
richer dynamics than the one of the theories constructed
\cite{Wess-grav} in terms of Moyal $\star$-products alone, by
twisting the algebra of diffeomorphisms with the frame-dependent
twist element (\ref{abelian twist}). The inconsistencies are caused
by the fact that the deformation of general coordinate
transformations is not so far done in a frame-independent manner. In
other words, when the twist element is chosen as (\ref{abelian
twist}), the frame-dependent Moyal $\star$-product is fixed once for
all by the choice of the twist and thus it does not transform at
all. Since the diffeomorphisms are basically external gauge
transformations, the situation is technically similar \cite{twist
gauge} to the one which results when one attempted to deform the
internal gauge transformations with the same twist element
(\ref{abelian twist}).

It thus appears that the currently studied noncommutative gauge and
gravitational theories show incompatibilities with respect to the
twisted \P symmetry, besides internal inconsistencies mentioned
above. It is therefore desirable to find a general symmetry
principle (and the applicability of the twisted \P symmetry leads to
the conclusion that this general symmetry will be a quantum one),
starting from which one could construct noncommutative gauge and
gravitational theories free of internal contradictions.

\section{Gauge transformations and the concept of twist}
\setcounter{equation}{0}

Let us consider the Lie algebra $\cal G$ as an internal symmetry.
The infinitesimal generators of the algebra $\cal G$ are denoted by
$T_a$, $a=1,...,m$ and they satisfy the commutation relations
\be\label{structure}[T_a,T_b]=if_{abc}T_c\,.\ee
Subsequently we gauge the algebra of internal symmetry $\cal G$ and
define
\be\label{gauge generator} \alpha(x)=\alpha^a(x)T_a\ee
as hermitian generators of the infinitesimal gauge transformations.
Since the gauge generators do not commute with the generators of the
global \P algebra, we can extend the \P algebra $\cal P$ by
semidirect product with the gauge generators, the purpose being to
eventually deform the enveloping algebra of this semidirect product,
$\cal U(\cal P\ltimes\cal G)$, considered as a Hopf algebra, by an
appropriately chosen twist \cite{Drinfeld-Reshetikhin} (see also
\cite{monographs}). The algebra of representation for $\cal U(\cal
P\ltimes\cal G)$ is the algebra of fields $\cal A$, defined on the
Minkowski space. The action of the generators of the Hopf algebra on
the fields is the usual one, even upon twisting. In particular, for
the infinitesimal gauge transformations we have
\be\label{gauge transf} \delta_\alpha \Phi(x)=i\alpha(x)\Phi(x)\,,\
\ \ \ \delta_\alpha \Phi^\dagger(x)=-i\Phi^\dagger(x)\alpha(x)\ee
where $\alpha(x)$ is defined in (\ref{gauge generator}) and
$\Phi(x)\in\cal A$. We emphasize the absence of a star-product in
(\ref{gauge transf}), unlike the case of the the traditional
noncommutative gauge theories \cite{Hayakawa}.

The principle of gauge invariance \cite{YM} requires the
introduction of gauge fields if we want the action of a theory to be
symmetric under local transformations. By their transformation
properties, the gauge fields have the role to compensate for the
terms arising from the fact that the derivatives of fields (in the
kinetic terms) do not transform according to the representations of
the {\it gauge} algebra, like the fields themselves do. With the
gauge fields $A_\mu(x)=A^a_\mu(x) T_a$ transforming in the adjoint
representation of the gauge algebra as
\be\delta_\alpha A_\mu(x)=i[\alpha(x),
A_\mu(x)]+\partial_\mu\alpha(x)\,,\ee
one constructs the covariant derivative
\be\label{covariant derivative}D_\mu=\partial_\mu-iA_\mu\,,\ee
such that the combination $D_\mu\Phi(x)$ transforms again like the
field itself under gauge transformations, i.e.
\be\label{covariant derivative transf} \delta_\alpha
D_\mu\Phi(x)=i\alpha(x)\left(D_\mu\Phi(x)\right)\,.\ee
Moreover, applying any number of covariant derivatives to a field,
the result will transform in the same way:
\be\label{many cov deriv transf} \delta_\alpha D_{\mu_1}\cdots
D_{\mu_n}\Phi(x)=i\alpha(x)\left(D_{\mu_1}\cdots
D_{\mu_n}\Phi(x)\right)\,,\ee
in other words,
\be\label{only cov deriv transf} \delta_\alpha D_{\mu_1}\cdots
D_{\mu_n}=[\alpha(x),D_{\mu_1}\cdots D_{\mu_n}]\,.\ee
We have to point out that, even upon twisting $\cal U(\cal
P\ltimes\cal G)$, the covariant derivative has to act as usual on
the matter fields, i.e. without any star-product between the gauge
field $A_\mu(x)$ and the matter field $\Phi(x)$. This is because the
covariant derivative is in effect a linear combination of generators
of $\cal U(\cal P\ltimes\cal G)$:
\be\label{covariant derivative}D_\mu=i(P_\mu-A_\mu^aT_a)\,,\ee
where the realization of $P_\mu$ on the Minkowski space,
$P_\mu=-i\partial_\mu$, is used.

\section{Non-Abelian twist of $\cal U(\cal P\ltimes\cal G)$}
\setcounter{equation}{0}

In \cite{twist gauge} it was shown in detail that the use of the
Abelian twist (\ref{abelian twist}) for deforming the Hopf algebra
$\cal U(\cal P\ltimes\cal G)$ is not compatible with the concept of
gauge transformations. We recall that the reason for this conflict
is the fact that the derivatives of a field do not transform
according to the representations of the gauge algebra, as the fields
themselves do.

However, the covariant derivatives of a field transform exactly
according to the same representation as the field itself, as we have
mentioned above. Thus the option of defining a {\it non-Abelian
twist element} involving covariant derivatives naturally occurs:
\be\label{nonabelian twist}{\cal
T}=\mbox{exp}\left(-{\frac{i}{2}\theta^{\mu\nu}D_\mu\otimes
D_\nu+{\cal O}(\theta^2)}\right)\,,\ee
where the terms of higher order in $\theta$ contain as well products
of covariant derivatives and remain to be found\footnote{For the
relaxation of the exponential form (\ref{nonabelian twist}) to an
arbitrary invertible function for the twist element $\cal T$, see
the end of this Section. The exponential form (\ref{nonabelian
twist}) however is taken, to start with, by requiring a
"correspondence principle", that the twist (\ref{nonabelian twist})
would reduce to the Abelian one (\ref{abelian twist}) in the absence
of gauge fields.}. The twist element (\ref{nonabelian twist}) has to
satisfy the twist conditions \cite{monographs}, i.e.:
\bea{\cal T}_{12}(\Delta_0\otimes id){\cal T}={\cal
T}_{23}(id\otimes \Delta_0){\cal T}\label{twist1}\\
(\epsilon\otimes id){\cal T}=1=(id\otimes\epsilon){\cal
T}\label{twist2}\eea
where $\Delta_0$ is the symmetric coproduct of the Lie algebra $\cal
P\ltimes\cal G$, such that
\be\label{symmetric coproduct}\Delta_0(Y)=Y\otimes 1+1\otimes Y\,,\
\ \mbox{for}\ \ Y\in\cal P\ltimes\cal G\,,\ee
$\epsilon:\cal U(\cal P\ltimes\cal G)\rightarrow{C}$ is the counit,
satisfying
\be(id\otimes\epsilon)\circ\Delta_0=id=(\epsilon\otimes
id)\circ\Delta_0\ee
and ${\cal T}_{12}={\cal T}\otimes1$ and ${\cal
T}_{23}=1\otimes{\cal T}$. By the twist element (\ref{nonabelian
twist}) one deforms the symmetric coproduct (\ref{symmetric
coproduct}):
\be\Delta_0(Y)\mapsto{\cal T}\Delta_0(Y){\cal T}^{-1}\,.\ee

The twisting of the coproduct of the generators requires a
corresponding deformation of the product of fields into a star
product, which we shall denote by $\bigstar$, to differentiate it
from the Weyl-Moyal $\star$-product:
\be\label{nonabelian twist prod}
m\circ(\Phi\otimes\Psi)=\Phi\Psi\rightarrow
m_\bigstar\circ(\Phi\otimes\Psi)=m\circ{\cal
T}^{-1}(\Phi\otimes\Psi)\equiv \Phi\bigstar\Psi\,.\nonumber\ee
Remark that the actual form of the covariant derivatives in the
second line of (\ref{nonabelian twist prod}) is given by the
respective fields on which they act, i.e. by their representation
under the gauge algebra. The associativity of the $\bigstar$-product
corresponding to the non-Abelian twist is equivalent to the
fulfilment of the twist condition (\ref{twist1}).

Since the purpose of the non-Abelian twist (\ref{nonabelian twist})
is to generalize the Abelian twist (\ref{abelian twist}), in a
manner which would consistently include the noncommutative gauge
transformations, the new star-product induced by the non-Abelian
twist has to reduce to the usual Weyl-Moyal star-product for
ordinary functions. Indeed, ordinary functions on the Minkowski
space have to be considered in the 1-dimensional (trivial)
representation of the gauge group $G$, i.e.
\be e^{i\alpha^a(x)T_a}\,f(x)=f(x)+i
\alpha^a(x)T_a\,f(x)+...=f(x)\,,\nonumber\ee
which implies $T_af(x)=0$. This means that for ordinary functions we
have $D_\mu f(x)=\partial_\mu f(x)$, from which it should follow:
\be\label{star_trivial} m\circ{\cal
T}^{-1}(f(x)\otimes g(x))\\
=m\circ\mbox{exp}\left({\frac{i}{2}\theta^{\mu\nu}\partial_\mu\otimes
\partial_\nu}\right)(f(x)\otimes g(x))\equiv
f(x)\star g(x)\,.\ee
The same result has to apply to the fields in the trivial
(1-dimensional) representation of the gauge group. It is then clear,
by taking $f(x)=x_\mu$ and $g(x)=x_\nu$ in the above, that the
non-Abelian twist would lead to gauge theories on the same
noncommutative space-time with the commutation relation (\ref{star
cr}). Therefore, in finding the concrete form of the non-Abelian
twist (\ref{nonabelian twist}) we have to fulfill the constraint
that the exponential has to reduce to the usual exponential function
of (\ref{abelian twist}) when its argument contains usual commuting
derivatives.

If we take in the non-Abelian twist (\ref{nonabelian twist}) only
the term of first order in $\theta$, one can straightforwardly show
that the twist condition (\ref{twist1}) is not fulfilled already in
the second order in $\theta$, while (\ref{twist2}) and
(\ref{star_trivial}) are. The second order terms which do not cancel
in (\ref{twist1}) are, in the l.h.s.
\be\label{lhs}
\frac{1}{2}\left(-\frac{i}{2}\right)^2\theta^{\mu\nu}\theta^{\rho\sigma}(D_\rho\otimes
D_\mu\otimes D_\sigma D_\nu+D_\mu\otimes D_\rho\otimes D_\sigma
D_\nu+2D_\mu D_\rho\otimes D_\nu\otimes D_\sigma+2D_\mu\otimes D_\nu
D_\rho\otimes D_\sigma) \ee
and in the r.h.s.
\be\label{rhs}
\frac{1}{2}\left(-\frac{i}{2}\right)^2\theta^{\mu\nu}\theta^{\rho\sigma}(2D_\rho\otimes
D_\mu\otimes D_\nu D_\sigma +D_\rho D_\mu \otimes D_\sigma\otimes
D_\nu+ D_\rho D_\mu \otimes D_\nu \otimes D_\sigma+2D_\rho\otimes
D_\mu D_\sigma\otimes D_\nu)\,. \ee

One may argue that there are still first order terms which were not
taken into account, i.e. $\theta^{\mu\nu}1\otimes F_{\mu\nu}$ and
$\theta^{\mu\nu}F_{\mu\nu}\otimes 1$. However, such terms will not
contribute to the cancelation of (\ref{lhs}) and (\ref{rhs}),
because it will introduce only terms in which the indices of the
second rank tensor, be it on the first, second or last place,
correspond to the same $\theta$, i.e. $F_{\mu\nu}\otimes
D_\rho\otimes D_\sigma$, $D_\rho\otimes F_{\mu\nu} \otimes D_\sigma$
or $ D_\rho\otimes D_\sigma\otimes F_{\mu\nu}$, while the indices of
the second rank tensor in the terms to be canceled of (\ref{lhs})
and (\ref{rhs}) correspond to different $\theta$s. Moreover, if one
writes an action with the new $\bigstar$-product replacing the usual
multiplication in the Lagrangean, the terms coming from
$\theta^{\mu\nu}1\otimes F_{\mu\nu}$ and
$\theta^{\mu\nu}F_{\mu\nu}\otimes 1$ will give the same
contribution, upon partial integration, like the terms coming from
$\theta^{\mu\nu}D_\mu\otimes D_\nu$. For this reasons we decide to
omit other terms of the first order in $\theta$ except
$\theta^{\mu\nu}D_\mu\otimes D_\nu$.

The second order terms not canceled in the twist condition suggest
the exponent in the form of a series in $\theta$. The general form
of such a series would be cumbersome to write down, however, we can
easily write the most general second order term which satisfies
(\ref{star_trivial}) and impose the twist condition (\ref{twist1})
up to second order in $\theta$.

Possible typical second order terms are:
\bea \theta^{\mu\nu}\theta^{\rho\sigma} (1\otimes D_\mu D_\nu D_\rho
D_\sigma) \ \ \mbox{and}\ \ \theta^{\mu\nu}\theta^{\rho\sigma}
(D_\mu D_\nu D_\rho D_\sigma\otimes 1)\label{zero-four}\\
\theta^{\mu\nu}\theta^{\rho\sigma} (D_\mu\otimes D_\nu D_\rho
D_\sigma) \ \ \mbox{and}\ \ \theta^{\mu\nu}\theta^{\rho\sigma}
(D_\mu D_\nu D_\rho\otimes D_\sigma)\label{one-three}\\
\theta^{\mu\nu}\theta^{\rho\sigma} (D_\mu D_\nu \otimes D_\rho
D_\sigma)\label{two-two}\,, \eea
with all the permutations of indices of the covariant derivatives.
The terms of the type (\ref{zero-four}) satisfy
(\ref{star_trivial}), but their structure is such, that they cannot
cancel the terms which appear in the second order from the first
term of the exponential. The terms of the type (\ref{two-two}) do
not satisfy in general (\ref{star_trivial}) but the terms which
satisfy the latter condition cannot help in the cancelation. The
only terms which satisfy (\ref{star_trivial}) and could contribute
to the cancelation are (\ref{one-three}) and we shall add only such
terms.

Since the order of the indices is important in the terms
(\ref{one-three}) with permutations, there are altogether
$2\frac{4!}{(4-3)!}=2\times 24$ of this type. However, due to the
antisymmetry of the indices $(\mu,\nu)$ and $(\rho,\sigma)$, only
$2\times 3$ combinations of indices are independent. Thus, the most
general form of (\ref{nonabelian twist}) with meaningful terms of
second order in $\theta$, which satisfy (\ref{star_trivial}), is:
\bea\label{nonabelian twist second order}{\cal
T}&=&\mbox{exp}\{(-\frac{i}{2}\theta^{\mu\nu}D_\mu\otimes D_\nu\cr
&+&\frac{1}{2}\left(-\frac{i}{2}\right)^2\theta^{\mu\nu}\theta^{\rho\sigma}[a\
D_\mu\otimes D_\sigma D_\nu D_\rho+b\ D_\mu\otimes D_\nu D_\sigma
D_\rho+c\ D_\mu\otimes D_\sigma D_\rho D_\nu\cr &+&a'\ D_\sigma
D_\nu D_\rho\otimes D_\mu+b'\ D_\nu D_\sigma D_\rho\otimes D_\mu+c'\
D_\sigma D_\rho D_\nu\otimes D_\mu+{\cal O}(\theta^2)]\}\,,\eea
where $a,b,c, a', b', c'$ are constants which have to be determined
by imposing (\ref{twist1}) up to the second order in $\theta$.
Typically, the terms which do not cancel out in the twist condition
are of the form $\theta\theta D\otimes D\otimes DD$, $\theta\theta
DD\otimes D\otimes D$ and $\theta\theta D\otimes DD\otimes D$.
Imposing the cancelation of the terms of the type $\theta\theta
D\otimes D\otimes DD$, one obtains $a=-1$ and $a+b+c=0$, while from
the terms of the type $\theta\theta DD\otimes D\otimes D$ one
obtains $a'=-1$ and $a'+b'+c'=0$. However, when requiring the
cancelation of the terms of the type $\theta\theta D\otimes
DD\otimes D$, one obtains $a+a'=2$ and $a+b+c=-(a'+b'+c')$.
Obviously the three conditions cannot be satisfied simultaneously,
consequently there are no second order terms, formulated in terms of
covariant derivatives, which can lead to the fulfillment of the
twist condition (\ref{twist1}) up to the second order in $\theta$.

Omitting the requirement that the non-Abelian twist should reduce to
the usual twist (i.e. with the usual Moyal $\star$-product) when
gauge fields are absent allows for other possible second order terms
terms, such as (\ref{two-two}), with all possible permutation of the
indices of covariant derivatives. We have verified that even by
admitting such terms, the twist condition (\ref{twist1}) cannot be
satisfied. We have also verified that, by relaxing the requirement
of exponential form for the twist as in (\ref{nonabelian twist}) to
an arbitrary invertible function $F(X)$, i.e. by taking the first
and second derivatives $F'(0)$ and $F''(0)$ (the coefficients of the
$\theta$-expansion of the twist) to be arbitrary, the twist
condition (\ref{twist1}) still cannot be satisfied. Thus the result
is general and is not based on the requirement of "correspondence
principle".

We can therefore conclude that a non-Abelian twist element, which
would generalize (\ref{abelian twist}) in a gauge covariant manner
cannot exist.

\section{Conclusions}

In this letter we have tackled the question whether the twist could
be regarded as a symmetry principle for the NC field and gauge
theories. To this end, we proposed a new, non-Abelian, twist element
(\ref{nonabelian twist}) for the formulation of noncommutative gauge
theories on Moyal spaces. The new star-product arising in this way,
containing covariant derivatives in place of the usual derivatives,
would insure both the twisted \P and the twisted gauge invariance of
noncommutative (gauge) field theories. We have shown, however, that
the non-Abelian twist element, although gauge covariant, does not
satisfy the twist conditions. The result does not depend on the
exponential form for the twist as in (\ref{nonabelian twist}), but
is valid for an arbitrary invertible functional form. Having in view
also the analysis of \cite{twist gauge}, which showed that the
Abelian twist (\ref{abelian twist}) cannot be used for twisting
gauge transformations, it appears that there is no way to reconcile
the twist condition and the gauge invariance principle. Let us
mention that by using the Seiberg-Witten map \cite{SW}, which
provides a connection between a NC gauge symmetry and the
corresponding commutative one as a power series in the
noncommutativity parameter $\theta_{\mu\nu}$, the resulting
Lagrangian or action \cite{madore} cannot be brought to the form
given by a twist.

It is intriguing that the {\it external} \P symmetry and the {\it
internal} gauge symmetry cannot be unified under a common twist. The
situation is reminiscent of the Coleman-Mandula theorem
\cite{coleman} (for a pedagogical presentation and other references,
see \cite{weinberg}), although not entirely, since the
Coleman-Mandula theorem concerns global symmetry and simple groups.
However, one can envisage that supersymmetry \cite{wess-zumino}, due
to its intrinsic internal symmetry, may reverse the situation, and a
noncommutative supersymmetric gauge theory can be constructed by
means of a twist \cite{progress}.

\vskip 0.5cm {\bf{Acknowledgements}}

We are indebted to M. Hayakawa, P. P. Kulish, K. Nishijima, P.
Pre\v{s}najder, M. Sheikh-Jabbari and A. A. Zheltukhin for useful
discussions.

\end{document}